# RAPID OSCILLATIONS IN CATACLYSMIC VARIABLES

# XVI. DW CANCRI

Joseph Patterson,[1] John R. Thorstensen,[2] Tonny Vanmunster,[3] Robert E. Fried,[4]

Brian Martin,[5] Tut Campbell,[6] Jeff Robertson,[6] Jonathan Kemp,[7,1]

David Messier,[8] & Eve Armstrong[1]

[1] Department of Astronomy, Columbia University, 550 West 120th Street, New York, NY 10027; jop@astro.columbia.edu, armstrong@astro.columbia.edu

[2] Department of Physics and Astronomy, Dartmouth College, 6127 Wilder Laboratory, Hanover, NH 03755; thorstensen@dartmouth.edu

[3] Center for Backyard Astrophysics (Belgium), Walhostraat 1A, B–3401 Landen, Belgium; Tonny.Vanmunster@cbabelgium.com

[4] Center for Backyard Astrophysics (Flagstaff), Braeside Observatory, Post Office Box 906, Flagstaff, AZ 86002; captain@asu.edu

[5] King's University College, Department of Physics, 9125 50th Street, Edmonton, AB T5H 2M1, Canada; bmartin@kingsu.ab.ca

[6] Arkansas Tech University, Department of Physical Science, 1701 North Boulder Avenue, Russellville, AR 72801; tutsky@yahoo.com, jeff.robertson@atu.edu

[7] Joint Astronomy Centre, University Park, 660 North A`ohōkū Place, Hilo, HI 96720; j.kemp@jach.hawaii.edu

[8] Center for Backyard Astrophysics (Norwich), 35 Sergeants Way, Lisbon, CT 06351; dpmessier@yahoo.com

**ABSTRACT**

We report photometry and spectroscopy of the novalike variable DW Cancri. The spectra show the usual broad H and He emission lines, with an excitation and continuum slope characteristic of a moderately high accretion rate. A radial-velocity search yields strong detections at two periods, 86.1015(3) min and 38.58377(6) min. We interpret these as respectively the orbital period $P_{orb}$ of the binary, and the spin period $P_{spin}$ of a magnetic white dwarf. The light curve also shows the spin period, plus an additional strong signal at 69.9133(10) min, which coincides with the difference frequency $1/P_{spin} - 1/P_{orb}$. These periods are stable over the 1 year baseline of measurement.

This triply-periodic structure mimics the behavior of several well-credentialed members of the "DQ Herculis" (intermediate polar) class of cataclysmic variables. DQ Her membership is also suggested by the mysteriously strong sideband signal (at $\nu_{spin}-\nu_{orb}$), attesting to a strong pulsed flux at X-ray/EUV/UV wavelengths. DW Cnc is a new member of this class, and would be an excellent target for extended observation at these wavelengths.

*Subject headings*: accretion, accretion disks — binaries: close — novae, cataclysmic variables — stars: individual (DW Cancri)

# 1. INTRODUCTION

DW Cancri is a little-studied cataclysmic variable, discovered in the Byurakan blue-galaxy Survey and identified as a CV from its Balmer emission lines (Stepanian 1982, Kopylov et al. 1988). Uemura et al. (2002) reported kilosecond quasi-periodic oscillations, centered on 37 and 73 minutes. Rodriguez-Gil et al. (2004, herafter RGAC) reported a radial-velocity period in the range 77–86 minutes, and suggested photometric periods near 38 and 77 min, although their data were severely plagued by aliases.

The curious light curves in the Uemura et al. study launched DW Cnc as a high-priority target for our global network of telescopes, the Center for Backyard Astrophysics (CBA; Skillman & Patterson 1993). We invested many nights in 2002–4 studying the star with CBA photometry, along with contemporaneous spectroscopy. In this paper we report the results, which certify DW Cnc as a new member of the DQ Herculis (intermediate polar) class of cataclysmic variable.

# 2. PHOTOMETRY

## 2.1 CAMPAIGN AND LIGHT CURVES

We obtained time-series photometry of DW Cnc for a total of 454 hours over 90 nights in 2002–4. About half the data was acquired through filters (*U*, *V*, *I*) on the MDM 1.3 m telescope, and about half at several CBA observing stations (with the lion's share from CBA–Belgium, CBA–Arkansas, CBA–Calgary, and Braeside Observatory). Some nights of CBA observation used a V filter, but most were unfiltered (to maximize signal-to-noise on this faint star). The primary comparison stars used were 1.9' E (*V*=15.21, *B–V*=0.68, *V–R*=0.39) and 2.5' WNW (*V*=12.70, *B–V*=0.46, *V–R*=0.30; magnitudes from Henden 2004) from DW Cnc. We found that aside from small and measurable zero-point shifts, the *V* and unfiltered time series were substantially the same, and could be spliced together for our main purpose, the study of periodic signals. The star also cooperated by remaining within ~0.1 mag of *V*=14.8 throughout our observations. This appears to be the normal "bright state" of DW Cnc. The photometric observing log is summarized in Table 1.

A typical night's light curve is shown in the upper frame of Figure 1. The light curve seems unremarkable for CVs, except that the apparent flickering is of unusually large amplitude on both long (~1 hour) and short (seconds) timescales. Power spectra of individual nights show no stable features at high frequency (out to 0.2 Hz), but consistent peaks near 37 and 21 c/d[9]. In fact, most of the variance at low frequency comes from the presence of these signals, which turn out to be very stable. The mean nightly power spectrum — the incoherent sum over the ten longest and best nights — is shown in the lower frame of Figure 1. The continuum power in this figure follows $P(\nu) \sim \nu^{-1.1}$, a rather flat distribution for a CV (exponents of −1.5 to −2.2 are the rule).

---

[9] In this paper we use the old-style "c/d" notation as a shorthand for cycles day$^{-1}$. Likewise, we frequently use a *truncated* HJD date, namely HJD = HJD − 2,452,000.

## *2.2 PERIODIC SIGNALS*

Since the campaign is long and the data spans a large range in terrestrial longitude, we are able to detect periodic signals with great sensitivity and (substantial) freedom from aliases. So we spliced the data in various convenient clusters, and analyzed for periodic content by calculating the power spectra of these clusters. The interval of densest coverage was in February 2003 (JD 671–93), and the power spectrum is shown in the upper frame of Figure 2. Very powerful signals are seen at 20.595 and 37.321(6) c/d, and another significant signal at 12.993(7) c/d. Each signal is essentially sinusoidal to within limits of measurement, since no harmonics are detected; indeed, there are no other periodic signals at all (to a semi-amplitude upper limit ~0.01 mag) out to 300 c/d. After "pre-whitening" the data by subtracting these three best-fitting sinusoids, we calculated the power spectrum of the residual time series. The result, seen in the lower frame of Figure 2, shows only a weak and broad bump near 16 c/d.

We repeated this analysis for five other clusters, with basically similar results. An example is the December 2002 cluster (JD 613–22), seen in Figure 3. After subtraction of the two obvious signals, we obtained the power spectrum seen in the lower frame. In the latter case, however, we "cleaned" the power spectrum by removing the alias pattern of the two candidate signals, at 12.99 and 16.80(2) c/d. These signals are too weak to be considered secure, but the one at lower frequency acquires some extra credibility by its agreement with the secure detection in Figure 2.

We calculated times of minimum light in each detected periodicity for each cluster of data. These timings are given in Table 2, and reduced to O–C diagrams in Figure 4 (where each point represents an average over 3–10 nights). The strong periodic signals track the following ephemerides:

$$\text{minimum light} = \text{HJD } 2452614.0240(7) + 0.0267942(3)\ E \qquad (1)$$
$$\text{minimum light} = \text{HJD } 2452614.0195(7) + 0.0485509(7)\ E. \qquad (2)$$

In Figure 5 we present mean waveforms for each of the detected signals during the 2003 February segment. They appear roughly sinusoidal, as suggested by the power spectra. The semi-amplitudes are 0.078(4), 0.084(4), 0.034(7), and 0.023(9) mag, for the four signals (respectively 70 min, 38 min, 110 min, and orbital — as we shall refer to these signals at 20.60, 37.32, 12.99, and 16.72 c/d.). Other segments gave similar results for the two strong signals. Thus it appears that these signals maintain essentially constant amplitude, although we know this only for long timescales (weeks). Amplitude change on shorter timescales would probably elude detection in this study.

Since the spectroscopy reported below appears to specify the true orbital period with high precision, we studied the data carefully for its possible presence. A marginal detection in the best cluster is possible (shown in Figure 5); but this detection weakened considerably when the entire year's data was considered. So it would be better to say that no *stable* orbital modulation is present, to a semiamplitude upper limit of 0.02 mag.

### 2.3 COLORS

Like nearly all CVs, DW Cnc is a very blue star, with $B-V \sim 0$, $U-B \sim -0.8$. We obtained 10 nights of multicolor photometry, in order to study the pulsed fraction in several passbands. The main signals, at 20.61 and 37.31 c/d, were found to have similar amplitudes in $V$ and $I$, but the 20.61 c/d signal was somewhat weaker in $U$ (0.05 compared to 0.08 mag semi-amplitude). The data were not strictly simultaneous, and there are night-to-night variations of up to 30% in pulse amplitude, so this constraint is not very strong. But to a first approximation, both pulse colors are fairly blue (similar to the star's mean color), with the 20 c/d signal somewhat less blue.

### 3. SPECTROSCOPY

Our spectroscopy is from four observing runs in 2002–3 (Table 3). We used the Hiltner 2.4 m telescope at MDM, the Modular spectrograph, and a thinned $2048^2$ Tektronix CCD yielding 2 Å/pixel from 4300 to 7500 Å. The signal-to-noise decreased at each end due to vignetting in the camera. The 1" slit gave a spectral resolution of 3.5 Å, and frequent exposure of arc lamps served to maintain an accurate wavelength scale. We observed flux standards and hot stars to allow conversion to absolute flux, and reduced to flux versus wavelength using standard procedures in IRAF. Individual exposures of DW Cnc were 4 to 6 minutes.

Figure 6 shows the mean spectrum from 2002 February. The continuum is quite blue, with $F_\lambda \propto \lambda^{-1.8}$ and emission lines of moderately high excitation. He II emission at $\lambda 4686$ and $\lambda 5411$ is present. Fluxes and equivalent widths are contained in Table 4. Double peaked emission is evident in the weaker He I lines and in Fe II $\lambda 5169$, with separations near ±460 km/s.

We measured radial velocities of H$\alpha$ by convolving the line with the derivative of a Gaussian, optimized for a 16 Å FWHM (Schneider & Young 1980); these velocities are given in Table 5. We searched for frequencies in the 0–20 c/d range using a "residualgram" method (Thorstensen & Freed 1985), and found two incommensurate frequencies, near 16.7 and 37.3 c/d. These signals and the resultant velocity curves are shown in the upper three frames of Figure 7. They appear to be coherent over the duration of the observations, and Monte Carlo tests show that there is no significant uncertainty in cycle count on a daily, monthly, or annual timescale.

To refine this further we fit the velocities with sinusoids $v(t) = \gamma + K\sin[2\pi(t-T_0)/P]$ at the two frequencies, then prepared two new time-series with the fitted sinusoid subtracted. Each was searched again, with the result seen in the lower frames of Figure 7. The alias structure ("picket fence") of the subtracted signal disappeared, and no other significant signals were unmasked. Folds of the new time-series gave a much cleaner result (lower right frames of Figure 7) and better-determined fit parameters (Table 6). We then simultaneously fit the original data with two sinusoids, with frequencies fixed at the values in Table 6. Figure 8 shows the data for our best six-night time-series, with the two-frequency fit superposed.

Figure 9 shows grayscale representations of the spectra averaged in phase, for both ephemerides; the technique is described by Taylor, Thorstensen, & Patterson (1999). At the lower frequency (left panels), *S*-waves appear in the Hα and He I λ5876 emission lines. (They are more subtly present in other lines, too.) Similar *S*-wave features appear in disk-accreting CVs, and are thought to arise from the hotspot where the mass-transfer stream strikes the outer disk. Their appearance here strongly suggests that 86 minutes is the true orbital period. The grayscale plots at the higher frequency (right panels) show velocity modulation in the envelope of the Hα and He I λ5876 emission lines, but no *S*-wave. There is also a striking modulation in the strength of the He II λ4686 line at this frequency.

## 4. DISCUSSION

### *4.1 PERIODS: THE MAGIC CREDENTIAL*

These data prove conclusively the existence of a stable signal at a period much shorter than $P_{orb}$. This is basically the defining characteristic of the DQ Herculis (intermediate polar) class of cataclysmic variables, in which the fast period is interpreted as the spin of an accreting magnetic white dwarf (Patterson 1994, Chapter 9 of Hellier 2000). In this case, that period is 38.6 minutes. Most of the other periods have a natural place in this picture: 86.1 min is $P_{orb}$, and 69.9 min is the low-frequency orbital sideband (corresponding to $\omega_{spin}-\omega_{orb}$). The latter arises from the frequency with which the rotating searchlight beam from the white dwarf sweeps across the secondary and other structures fixed in the orbital frame.

Our description of the variability appears to be roughly consistent with previous photometry of DW Cnc. Uemura et al. (2002) and RGAC had smaller photometric data sets (respectively 75 and 11 hours, with an unfavorable distribution in time). Neither were well-suited for period study, and we suspect that this is why we reached different conclusions. When we simulated their data by selecting subsets of our data with their time distribution, we could (sometimes) produce power spectra resembling Figure 6 of Uemura et al. or Figure 3 of RGAC[10]. In detail, these three studies are not in particularly good agreement: Uemura et al. suggested that the signals are quasiperiodic, and RGAC reported signals at 38.51(2) and 77.37(3) min.

The spectroscopy of RGAC is more obviously consistent with ours. In particular, they identify the critical feature of two noncommensurate frequencies, although their data are too heavily aliased to specify those frequencies uniquely. The present data establish this description with long-term ephemerides.

### *4.2 PHASING OF THE PERIODIC SIGNALS*

The basic geometry of the binary is usually quite poorly known in intermediate polars. The secondaries are generally invisible; and the "disk" emission lines are likely to be poor tracers of binary motion since they are heavily contaminated by Dopper motions in the two columns of infalling gas (the mass-transfer stream and the magnetically-channeled accretion column onto the white dwarf). This is true in DW Cnc also; our ephemeris accurately describes the motion of an emission line, but we do not know what that implies about the true dynamical motions of the stars.

---

[10] RGAC report that the signal is "remarkably coherent" over 4 years; but that was probably a misprint, since their power spectrum spans only 4 days. And indeed, that data cannot substantiate coherence over 4 days, since their Figure 3 shows dozens of acceptable periods. Effectively, the RGAC photometry tests coherence only for a single 6-hour observation (i.e. just a few cycles).

The geometry of the spin pulse can also be complex. Hellier (1997) summarizes the available data. Of the five IPs with $v_{rad}$ variations at $P_{spin}$, two (AO Psc and FO Aqr) show maximum blueshift at maximum light, as expected in the simplest interpretation of this component (arising from infalling gas onto the "back pole"; see Kim & Beuermann 1995). The other three (EX Hya, BG CMi, PQ Gem) fail to obey this simple relation. DW Cnc appears to join the miscreants in this respect, showing maximum blueshift 0.27±0.05 cycles before maximum light.

Thus, without additional assumptions which we are unwilling to make, the observed phase information does not yet constrain the binary and accretion geometries. A secure measurement of true orbital motion would help a great deal, as would a detection of the spin pulse in X-rays.

### 4.3 DISTANCE AND LUMINOSITY

Little information is available concerning the star's distance. Cataclysmic variables tend to follow an empirical law relating $M_V$ to $P_{orb}$ (Figures 5 and 7 of Patterson 1984), but DW Cnc is unusual because it is magnetic. It is probably wiser to compare with the one star it most closely resembles: EX Hya, the only other short-period magnetic CV with a long $P_{spin}$. A recent HST parallax puts EX Hya at 65 pc, implying $M_V$=8.8 (Beuermann et al. 2003). Since DW Cnc is another magnetic CV of short $P_{orb}$ and long $P_{spin}$, we might expect a similar $M_V$, suggesting a distance ~200 pc. Assuming a "bolometric" correction appropriate to its color (~1.2 mag), the 0.1–1.0 μ luminosity in the high state would be $4\times10^{32}$ erg/s $d_{200}^2$, where $d_{200}$ is its distance in units of 200 pc.

### 4.4 He II EMISSION, AND THE X–RAY SEARCHLIGHT

DW Cnc shows fairly strong He II λ4686 emission (EW~9 Å), which is sometimes used as an identifier of magnetic CVs. But caution is needed on this point. He II emission is rare in dwarf novae, and common in novalike variables (say at the level of EW~3–8 Å, see Figure 5 of Patterson & Raymond 1985). Stronger emission (say EW>10 Å) is generally limited to magnetic CVs and stars of very high $\dot{M}$ (the supersofts). It's worth exploring why. The origin of the line is very likely photoionization, and thus it mainly testifies to an abundance of X-ray/EUV photons (with energy above 54 eV). This implies a high-temperature region ($T>10^5$ K). Now in the blackbody approximation, $T$ scales as $(L/f)^{1/4}$, where $L$ is the accretion luminosity and $f$ is the fraction of white-dwarf area that is radiating. So high $T$ can be achieved with high luminosity (high $\dot{M}$) or with a very small $f$. Supersofts choose the former route. A small $f$ is most readily achieved with a magnetic field which channels accretion flow to a very small area around the magnetic pole. This is the main reason[11] that magnetic CVs consistently sport strong He II emission even when $L$ and $\dot{M}$ are quite modest.

But with EW(λ4686)~9 Å, and with Hβ emission ~10 times stronger, DW Cnc is only of moderately high excitation, not quite in the realm of the well-established magnetic CVs. The much more salient credential is the stable fast period. And the presence of a large-amplitude signal at $\omega_{spin}-\omega_{orb}$ (i.e. the 70 minute period) is perhaps the most compelling credential of all.

Why? The reason is that the 70 min signal is of large amplitude, nearly as strong as the 38 min signal which presumably drives it. That can only mean that the 38 min signal is very powerful indeed. The sideband signal must arise from reprocessing in nonaxisymmetric structures which orbit the white dwarf at exactly $P_{orb}$. The latter are most simply described as "the secondary", so let us adopt that assumption. Now in Roche geometry the secondary subtends a solid angle of ~3% of 4π sr at the white dwarf, so only a few percent of white-dwarf luminosity can be reprocessed in the secondary. This inefficiency is basically why "reflection effects" in the secondary are only rarely seen in CV light curves.

Thus sideband signals are generally expected to be quite weak, contrary to observation in DW Cnc. There are a few rescue strategies which could be considered.

(1) Beaming. Maybe the white-dwarf radiation illuminates the secondary powerfully, but illuminates terrestrial observers quite weakly.

(2) Maybe the absorbing fraction is much higher than 3%. Another discrete structure which orbits with $P_{orb}$ can be the mass-transfer stream, or more specifically the "hot spot" where the stream impacts the accretion disk. The hot-spot size is not well constrained, but could possibly subtend as much as 15% of 4π sr.

(3) Maybe the bolometric luminosity of the white dwarf's 38 minute signal is very high, due to the presence of hitherto unseen EUV and X-ray flux.

---

[11] Not the only reason. Hard X-rays, usually present in equal or greater supply, also contribute; but they are relatively inefficient ionizers of helium (since they waste much more energy per photon).

Each of these excuses could play a role. There is not much to say about (1), since none of the relevant angles are known for this binary. (2) likely plays some role, but even 15% is still inadequate, as we will see below. The dominant explanation is likely to be (3). And this is very plausible, since accretion onto a magnetic white dwarf is predominantly radial, and high temperatures are generally achieved in radial accretion flow. There must be a powerful UV/EUV/X-ray searchlight associated with that 38 minute signal.

More specifically, a minimum luminosity in the 70 minute signal can be estimated by assuming a pulsed fraction of 100%, and a bolometric correction of zero. In this case we obtain $L_{70} = 2 \times 10^{31} \, d_{200}^2$ erg/s. More plausibly, let's assume that the 70 minute signal is actually 50% pulsed[12], with a bolometric correction near 0.5 mag. This implies $L_{70} \sim 6 \times 10^{31} \, d_{200}^2$ erg/s. In order to power this with a prograde searchlight at $P_{spin}$=38 min, with a generous assumption of 15% reprocessing, we probably need a searchlight ~7 times stronger, or $L_{38} \sim 4 \times 10^{32} \, d_{200}^2$ erg/s.

Hard X-rays are a plausible choice for the powering searchlight. The only X-ray observation is in the *ROSAT* all-sky Survey, where the star appears to be detected at 0.34±0.12 counts/s. For an unabsorbed 10 keV spectrum, typical of CVs in general, this corresponds to a 1–10 keV flux of $9 \times 10^{-12}$ erg/cm$^2$/s, or $L_X = 5 \times 10^{31} \, d_{200}^2$ erg/s. Intermediate polars commonly show a fairly large absorption from cool gas near the emission site; assuming $N_H = 2 \times 10^{21}$ cm$^{-2}$, we obtain $L_X(1–10 \text{ keV}) = 10^{32} \, d_{200}^2$ erg/s. This is somewhat of an upper limit; ignoring uncertainties of source variability, a hard X-ray source much brighter than this would have been detected by the all-sky surveys of the 1970s (HEAO–1, HEAO–2).

The other strong candidate for the searchlight is the $L_{38}$ component in the ultraviolet. Again assuming a pulsed fraction of 50%, but with a bolometric correction of 1.5 mag (since the $L_{38}$ component appears to be slightly bluer than the star's mean color, whereas the $L_{70}$ component is slightly redder), we obtain $L_{38} = 1.5 \times 10^{32} \, d_{200}^2$ erg/s. Thus we get a likely optical/UV/X-ray $L_{38}$ component of $2 \times 10^{32} \, d_{200}^2$ erg/s — about half of what we need.

This shortfall can easily be hidden in the several uncertainties of bolometric correction and geometry, described above. Even so, we needed to assume that 15% of the searchlight falls on the "secondary" (where we use this term to denote "mass-donor-plus-hot-spot"), which could be considered optimistic. It is obvious that the $L_{38}$ searchlight powers the $L_{70}$ signal, but we do not yet understand the energetics of this.[13]

## 5. SUMMARY

---

[12] We do not know the true pulse fraction. It must be at least 16%, because that is observed. And it is probably not very close to 100%, because that is hard to achieve with a highly sinusoidal waveform. So ~50% is a plausible and simple choice.

[13] An obvious candidate is a strong EUV component, seen in a few intermediate polars (Haberl & Motch 1995). These are normally very weak in CVs, but stars with magnetic channeling can easily have small *f* and therefore high *T* — high enough to ionize helium and thereby avoid the edges that normally quench the emission. This is definitely worth a search.

1. We establish strictly periodic signals at 38.6 and 86.1 min in the radial velocities of DW Cnc. The presence of stable noncommensurate frequencies is a strong credential — almost a *defining* credential — for membership in the DQ Her (intermediate polar) class of cataclysmic variables.

2. Time-series photometry reveals strictly periodic 38.6 and 69.9 minute periodic signals, which maintain a constant phase over the 1 year baseline of measurement. The three frequencies follow a simple relation (1/69.9) = (1/38.6)–(1/86.1), to within limits of measurement. This is a well-known hallmark of the DQ Her class, and suggests that 86.1 min should be regarded as $P_{orb}$, 38.6 min as $P_{spin}$, and 69.9 min as the "reprocessing" period (lower orbital sideband). Ephemerides for all three signals have an expiration date around 2006.

3. The existence of so strong a signal (0.16 mag full amplitude) at 70 min implies a strong "searchlight" signal at 38 min sweeping over the parts of the binary fixed in the orbital frame (secondary, mass-transfer stream). This is likely to be powered in part by a pulsed hard X-ray source (with $L_X \sim 10^{32} d_{200}^2$ erg/s), but most of the searchlight's power probably appears at UV and EUV wavelengths. Since all of these bandpasses are unexplored, searches for this component are very desirable!

4. There is also some evidence for a weak periodic signal at 110.85(9) min. This does not seem related to any of the other clocks in the binary, and we leave it as an unsolved problem.

We acknowledge financial support from NSF (AST–00–98254) and NASA (NNG04GA83G), and thank Jerry Foote, Thomas Krajci, Donn Starkey, and David Skillman for contributions to the photometry campaign. We dedicate this paper to our dear friend and colleague Bob Fried, who died on 13 Nov 2003 while piloting his plane to a remote airstrip in Black Canyon, AZ, on one of his many "Angel Flight" trips to transport needy people for medical help. Bob was one of the most charismatic amateur astronomers in the world, and a legend in the world of variable stars. This paper was one of his last major projects, although his data were widely available to the astronomical community, and will surely be appearing in scientific journals for many years to come.

TABLE 1
2002–4 PHOTOMETRY

| Telescope | Observer | Nights/hours |
|---|---|---|
| MDM 1.3 m | E. Armstrong, J. Kemp | 32/167 |
| Braeside 0.4m | R. Fried | 14/75 |
| CBA–Belgium 0.35 m | T. Vanmunster | 13/63 |
| CBA–Arkansas 0.4 m | T. Campbell, J. Robertson | 11/49 |
| CBA–Norwich 0.25 m | D. Messier | 8/46 |
| CBA–Alberta 0.3 m | B. Martin | 6/33 |
| Others | | 6/25 |

TABLE 2
PHOTOMETRIC MINIMA
(HJD 2,452,000+)

| 0.0485 d signal | 0.0268 d signal |
|---|---|
| 614.0195 | 614.0232 |
| 635.9671 | 635.9699 |
| 671.6476 | 671.6584 |
| 701.7025 | 701.6677 |
| 723.6925 | 723.6919 |
| 994.6107 | 994.6090 |
| 1007.7190 | 1003.7465 |
|  | 1007.7104 |

NOTE: Each point is an average over 3–10 nights.

Table 3. Journal of Spectroscopy

| UT Date | N | First HA | Last HA |
|---|---|---|---|
| 2002 Jan 20 | 2 | $+4:06$ | $+4:10$ |
| 2002 Jan 21 | 1 | $+3:35$ | $+3:35$ |
| 2002 Jan 23 | 2 | $-1:08$ | $-0:27$ |
| 2002 Jan 24 | 39 | $-3:27$ | $+4:34$ |
| 2002 Feb 16 | 9 | $-0:48$ | $+2:31$ |
| 2002 Feb 19 | 20 | $-2:16$ | $+3:03$ |
| 2002 Feb 21 | 17 | $-0:08$ | $+1:34$ |
| 2002 Dec 12 | 45 | $-3:50$ | $+2:03$ |
| 2002 Dec 13 | 25 | $-3:21$ | $-1:09$ |
| 2002 Dec 14 | 24 | $-4:19$ | $+2:57$ |
| 2002 Dec 19 | 31 | $-2:59$ | $+2:58$ |
| 2003 Feb 02 | 18 | $+1:51$ | $+3:56$ |

Table 4. Emission Lines

| Feature | E.W.[a] (Å) | Flux ($10^{-15}$ erg cm$^{-2}$ s$^{-1}$) | FWHM (Å) |
|---|---|---|---|
| H$\gamma$ | 71 | 532 | 25 |
| HeI $\lambda$4471 | 15 | 107 | 22 |
| HeI $\lambda$4686 | 9 | 60 | 21 |
| H$\beta$ | 80 | 475 | 26 |
| HeI $\lambda$4921 | 6 | 39 | 22 |
| HeI $\lambda$5015 | 7 | 37 | 24 |
| Fe $\lambda$5169 | 4 | 19 | 27 |
| HeI $\lambda$5876 | 21 | 90 | 25 |
| H$\alpha$ | 91 | 321 | 26 |
| HeI $\lambda$6678 | 11 | 36 | 28 |
| HeI $\lambda$7067 | 8 | 24 | 30 |

[a]Emission is counted as positive.

Table 5. H$\alpha$ Radial Velocities

| HJD[a] | $V$[b] | HJD[a] | $V$[b] | HJD[a] | $V$[b] | HJD[a] | $V$[b] | HJD[a] | $V$[b] | HJD[a] | $V$[b] |
|---|---|---|---|---|---|---|---|---|---|---|---|
| 294.9866 | 48 | 298.9816 | −51 | 326.7443 | −20 | 620.8992 | −28 | 621.8613 | 76 | 627.8321 | −43 |
| 294.9893 | 25 | 298.9849 | −30 | 326.7483 | 30 | 620.9025 | 72 | 621.8647 | 111 | 627.8362 | 19 |
| 295.9627 | 76 | 298.9883 | 39 | 326.7523 | 104 | 620.9059 | 74 | 621.8681 | 134 | 627.8402 | 110 |
| 297.7608 | 73 | 298.9916 | 105 | 326.7564 | 125 | 620.9113 | 22 | 621.8714 | 123 | 627.8443 | 124 |
| 297.7894 | 74 | 298.9950 | 141 | 326.7604 | 47 | 620.9147 | −20 | 622.7373 | −67 | 627.8483 | 91 |
| 298.6620 | −25 | 321.7089 | 5 | 326.7668 | −45 | 620.9181 | 16 | 622.7430 | −65 | 627.8546 | 36 |
| 298.6675 | −6 | 321.7144 | 106 | 326.7708 | −51 | 620.9214 | 74 | 622.7464 | −26 | 627.8586 | 62 |
| 298.6708 | 8 | 321.7198 | 124 | 326.7749 | −50 | 620.9248 | 112 | 622.7498 | −22 | 627.8627 | 121 |
| 298.6741 | 9 | 321.7265 | 128 | 326.7789 | 6 | 620.9726 | 4 | 622.7531 | 6 | 627.8668 | 176 |
| 298.6775 | −10 | 321.8233 | 42 | 326.7829 | 53 | 620.9760 | 58 | 622.7565 | 11 | 627.8708 | 168 |
| 298.6808 | −42 | 321.8301 | 66 | 326.7892 | 9 | 620.9793 | 112 | 622.7620 | −34 | 627.9931 | 12 |
| 298.6841 | −48 | 321.8349 | −24 | 326.7932 | −40 | 620.9827 | 126 | 622.7653 | −22 | 627.9965 | 40 |
| 298.6875 | −58 | 321.8396 | 54 | 620.7627 | 43 | 620.9861 | 122 | 622.7687 | 53 | 627.9999 | 66 |
| 298.6908 | −32 | 321.8467 | 70 | 620.7660 | 55 | 620.9937 | 33 | 622.7721 | 110 | 628.0032 | 34 |
| 298.8031 | 76 | 324.6392 | 74 | 620.7694 | 24 | 620.9971 | −15 | 622.7754 | 141 | 628.0066 | −26 |
| 298.8086 | 153 | 324.6432 | 68 | 620.7728 | −53 | 621.0004 | −17 | 622.9946 | 75 | 628.0130 | −85 |
| 298.8119 | 141 | 324.6472 | 24 | 620.7762 | −100 | 621.0038 | −16 | 623.0000 | −24 | 628.0164 | −69 |
| 298.8153 | 88 | 324.6513 | 4 | 620.7815 | −92 | 621.0072 | 10 | 623.0034 | −41 | 628.0197 | −42 |
| 298.8186 | 83 | 324.6553 | 19 | 620.7849 | −82 | 621.7801 | −15 | 623.0068 | −15 | 628.0231 | 10 |
| 298.8220 | 44 | 324.6617 | 118 | 620.7882 | 1 | 621.7835 | −48 | 623.0101 | 33 | 628.0265 | 78 |
| 298.8253 | 32 | 324.6658 | 73 | 620.7916 | 81 | 621.7868 | −42 | 623.0135 | 132 | 672.8580 | 27 |
| 298.8287 | 79 | 324.6698 | −6 | 620.7950 | 106 | 621.7902 | −67 | 623.0168 | 175 | 672.8637 | −51 |
| 298.8320 | 109 | 324.6738 | −97 | 620.8004 | 72 | 621.7936 | −121 | 623.0224 | 142 | 672.8670 | −71 |
| 298.8354 | 84 | 324.6779 | −111 | 620.8038 | 54 | 621.7989 | −83 | 623.0258 | 29 | 672.8704 | −20 |
| 298.8415 | −33 | 324.8213 | 96 | 620.8072 | −0 | 621.8022 | −63 | 623.0291 | −29 | 672.8738 | 30 |
| 298.8449 | −88 | 324.8254 | 169 | 620.8105 | 25 | 621.8056 | 11 | 623.0325 | −34 | 672.8771 | 133 |
| 298.8482 | −110 | 324.8294 | 177 | 620.8139 | 35 | 621.8090 | 80 | 623.0359 | −12 | 672.8805 | 171 |
| 298.8515 | −95 | 324.8335 | 100 | 620.8201 | 101 | 621.8123 | 130 | 623.0392 | −4 | 672.8839 | 205 |
| 298.8549 | −60 | 324.8379 | 30 | 620.8235 | 67 | 621.8178 | 127 | 627.7795 | −37 | 672.8872 | 186 |
| 298.8582 | −3 | 324.8439 | 10 | 620.8268 | 37 | 621.8211 | 102 | 627.7869 | 98 | 672.9178 | −141 |
| 298.8616 | 45 | 324.8480 | 48 | 620.8302 | −17 | 621.8245 | 54 | 627.7910 | 113 | 672.9212 | −120 |
| 298.8649 | 65 | 324.8520 | 65 | 620.8336 | −77 | 621.8279 | 41 | 627.7950 | 84 | 672.9245 | −82 |

Table 5—Continued

| HJD[a] | V[b] | HJD[a] | V[b] | HJD[a] | V[b] | HJD[a] | V[b] | HJD[a] | V[b] | HJD[a] | V[b] |
|---|---|---|---|---|---|---|---|---|---|---|---|
| 298.8682 | 61 | 324.8560 | 5 | 620.8732 | 142 | 621.8313 | 38 | 627.7991 | 42 | 672.9279 | −42 |
| 298.8716 | 34 | 324.8600 | −86 | 620.8766 | 145 | 621.8391 | 125 | 627.8031 | 6 | 672.9313 | 95 |
| 298.9092 | −25 | 326.7221 | −18 | 620.8799 | 101 | 621.8425 | 99 | 627.8093 | 81 | 672.9346 | 150 |
| 298.9682 | −2 | 326.7261 | 42 | 620.8833 | −57 | 621.8459 | 24 | 627.8134 | 64 | 672.9380 | 178 |
| 298.9716 | 4 | 326.7302 | 111 | 620.8867 | −97 | 621.8492 | −76 | 627.8174 | 94 | 672.9414 | 160 |
| 298.9749 | −54 | 326.7342 | 58 | 620.8924 | −122 | 621.8526 | −62 | 627.8215 | 35 | 672.9447 | 114 |
| 298.9783 | −71 | 326.7382 | 2 | 620.8958 | −88 | 621.8580 | 43 | 627.8256 | −37 | | |

[a]Heliocentric JD of mid-integration minus 2452000.

[b]Heliocentric H$\alpha$ velocity (km s$^{-1}$).

Table 6. Sine Fits

| Fit | $T_0$[a] | $P$ (d) | $K$ (km s$^{-1}$) | $\gamma$ (km s$^{-1}$) | $\sigma$ (km s$^{-1}$) |
|---|---|---|---|---|---|
| Long $P$ | 52620.7910(9) | 0.0597929(3) | 82(8) | 32(5) | 55 |
| Short $P$ | 52620.7614(9) | 0.0267942(3) | 78(8) | 33(5) | 57 |
| Long; short $P$ subtr. | 52620.7902(6) | 0.05979267(18) | 69(4) | $\cdots$ | 29 |
| Short; long $P$ subtr. | 52620.7618(3) | 0.02679429(4) | 66(4) | $\cdots$ | 29 |
| Long $P$ (simult.) | 52620.7902(6) | [0.05979267] | 72(4) | 28(3) | 30 |
| Short $P$ (simult.) | 52620.7618(3) | [0.02679429] | 67(4) | $\cdots$ | $\cdots$ |

[a]Heliocentric JD minus 2 400 000.

Note. — Parameters of least-squares sinusoids of the form $v(t) = \gamma + K \sin[2\pi(t-T_0)/P]$, fitted to the 233 H$\alpha$ velocities. The last two lines are from a fit in which sinusoids at the two frequencies were fitted simultaneously, so $\gamma$ and $\sigma$ are common to the two fits.

# FIGURE CAPTIONS

FIGURE 1. — *Upper frame*, sample light curve of DW Cnc. *Lower frame*, the nightly average power spectrum, the incoherent sum over the 10 longest nights.

FIGURE 2. — *Upper frame*, power spectrum of a 23-night light curve in 2003, showing three signals with their frequencies (±0.006 c/d) flagged by arrows. *Lower frame*, power spectrum of the residuals, after these signals are removed. Only a possible very weak signal near 16 c/d remains; the arrow points to the "orbital" frequency identified by spectroscopy.

FIGURE 3. — *Upper frame*, power spectrum of a 10-night light curve in December 2002, showing strong signals consistent with those of Figure 2. *Lower frame*, power spectrum of the residuals, after removing the strong signals. Probable detections at 12.99 and 16.80 (±0.02) c/d.

FIGURE 4. — O–C diagrams of the timings of photometric minima, relative to Eqs. (1) and (2). The signals maintain constant phase over the 400-day baseline.

FIGURE 5. — Mean waveforms in *V* light of four photometric signals during 2003 February. HJD times of maximum light and semi-amplitudes were as follows: 38 min, 671.6444, 0.084 mag; 70 min, 671.6750, 0.078 mag; 110 min, 671.6624, 0.034 mag; orbital, 671.6537, 0.023 mag.

FIGURE 6. — Mean spectrum of DW Cnc from 2002 February. Line identifications and strengths are given in Table 4.

FIGURE 7. — Period searches of Hα emission velocities. The upper three frames show the raw "residualgram", and folded velocity curves for the two obvious signals. The lower four frames show the same after subtraction of the contaminating signals.

FIGURE 8. — Hα radial velocities from the six nights of longest coverage. Each frame covers the same time interval.

FIGURE 9. — Montage of grayscale representations of phase-averaged spectra. The vertical axes show phase increasing vertically, and the horizontal axes are wavelength in Å. Left and right panels are respectively folded on the long (86.1 min) and short (38.6 min) period. The top panels show the region around Hβ, He II λ4686, He I λλ4922 and 5015, and Fe I λ5169; the middle panels show He I λ5876; the bottom panels show Hα. Grayscales are selected to maximize visibility of the periodic features.

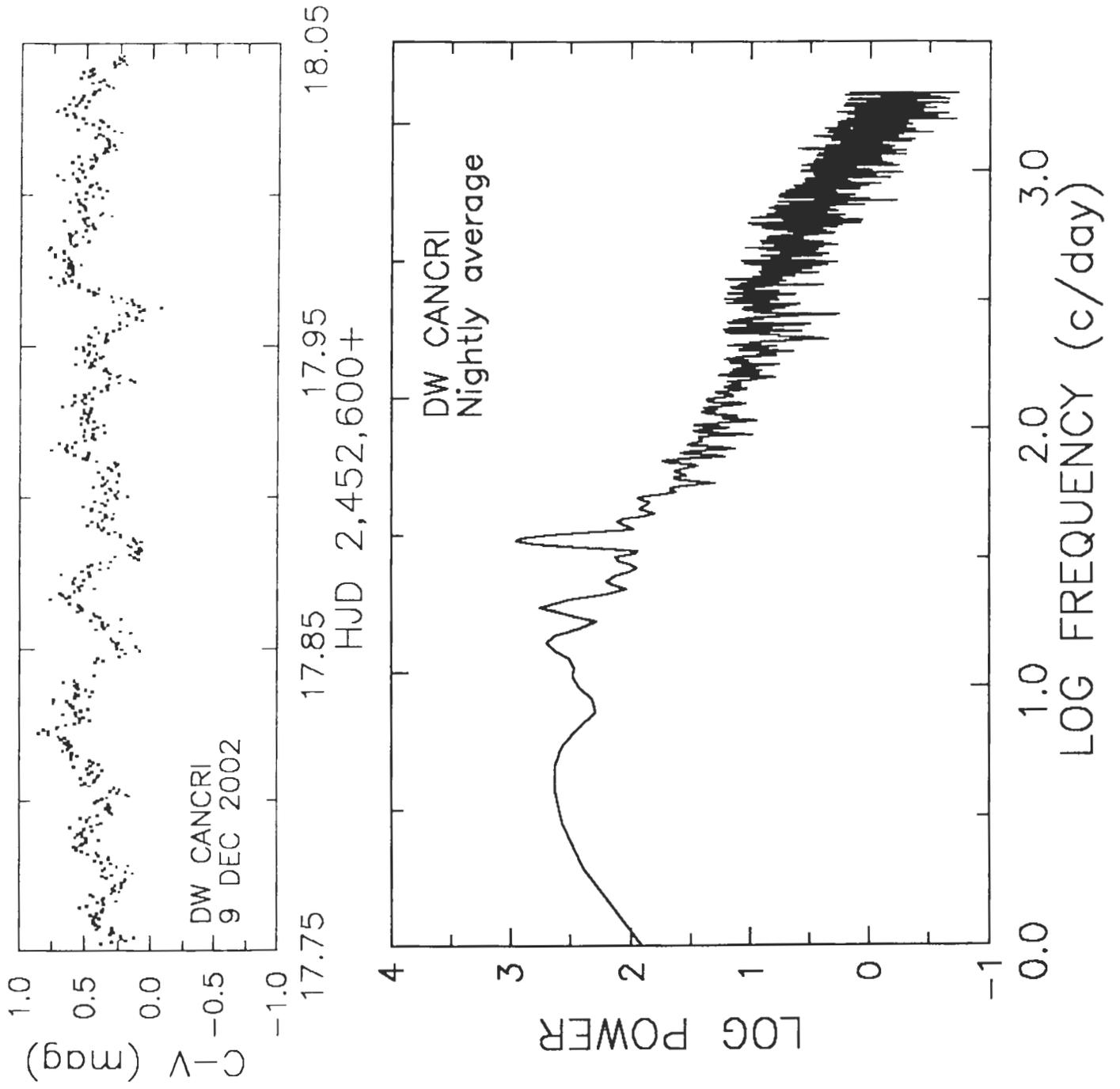

Fig. 1

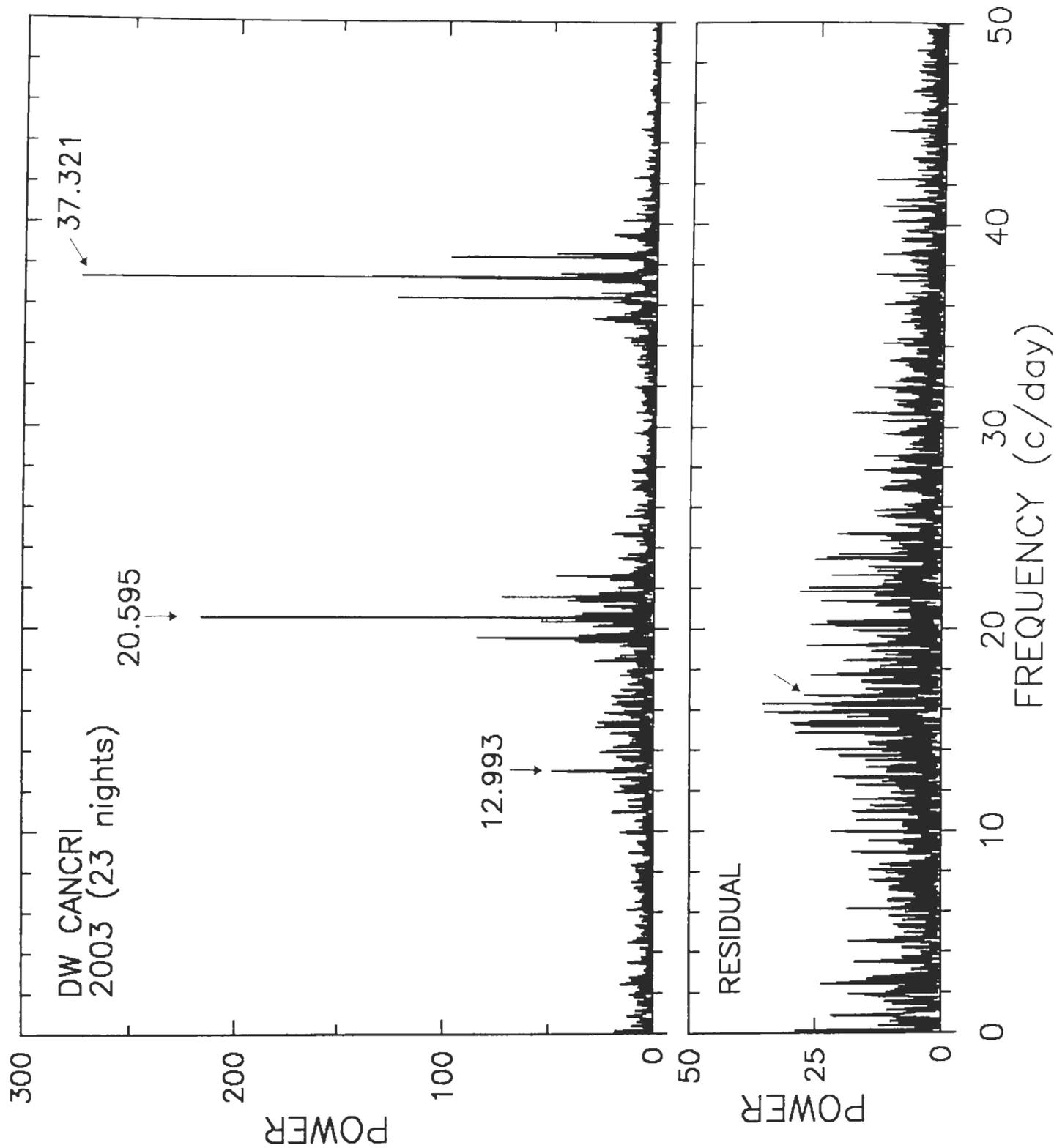

Fig 2

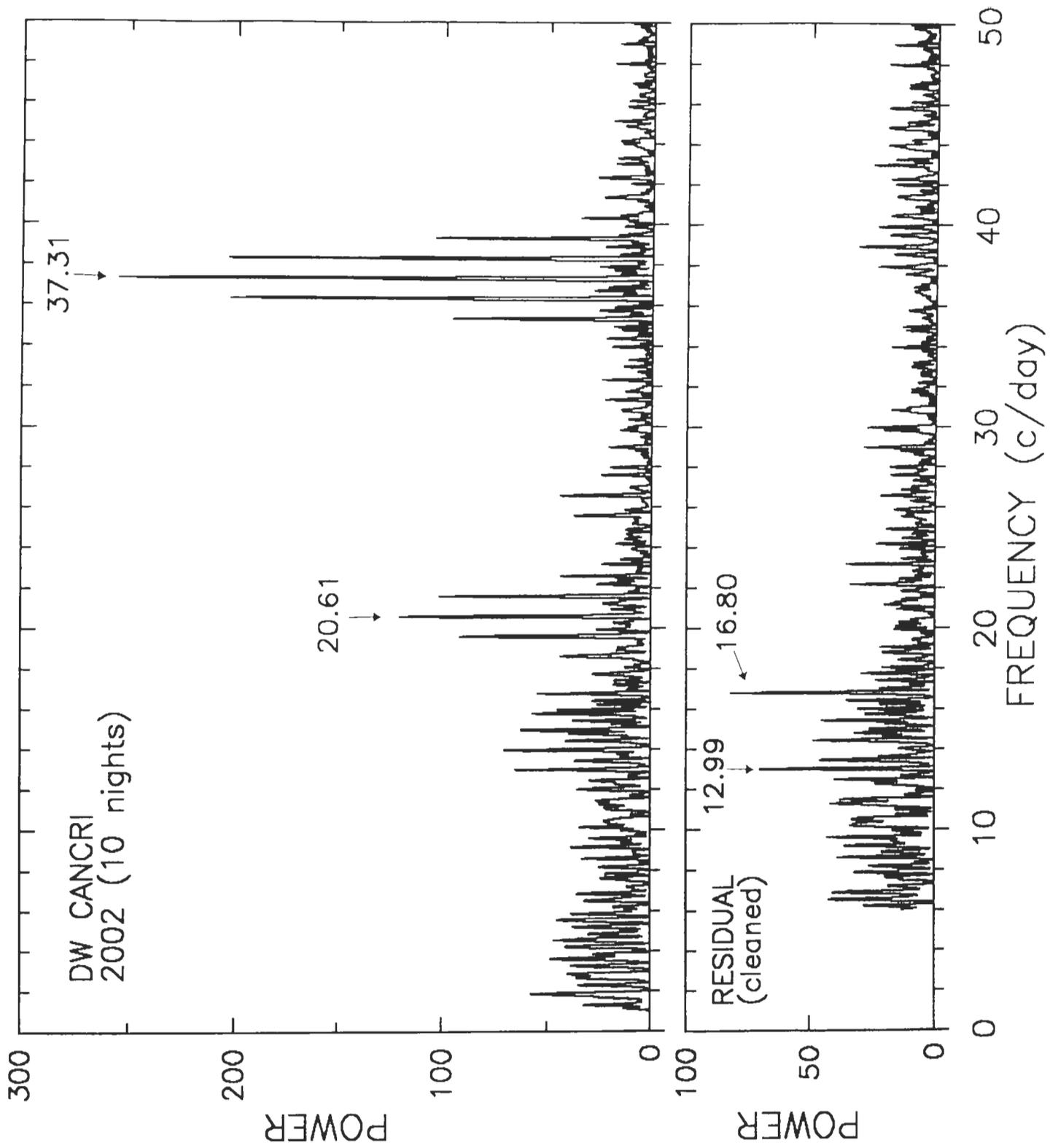

Fig 3

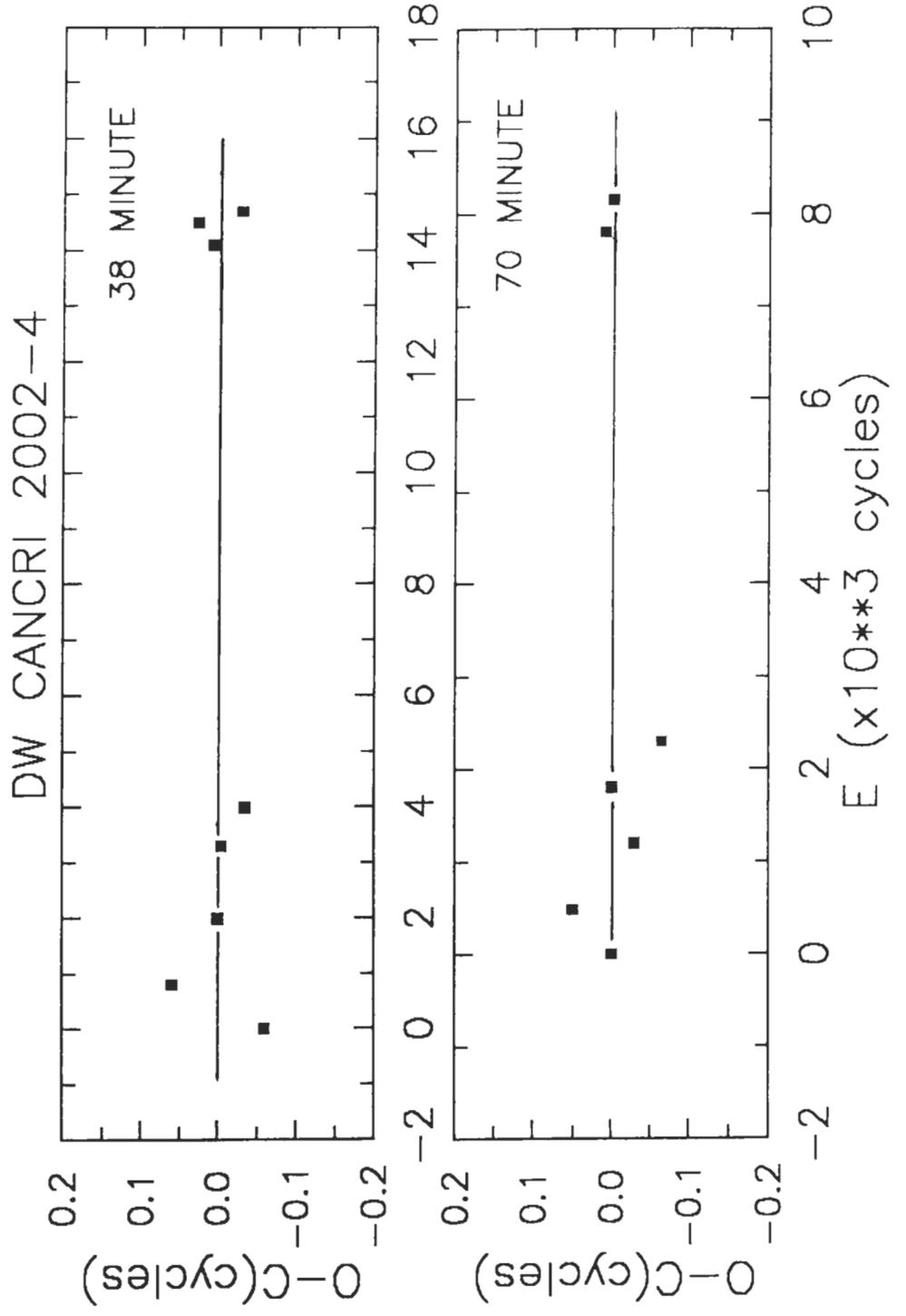

Fig. 4

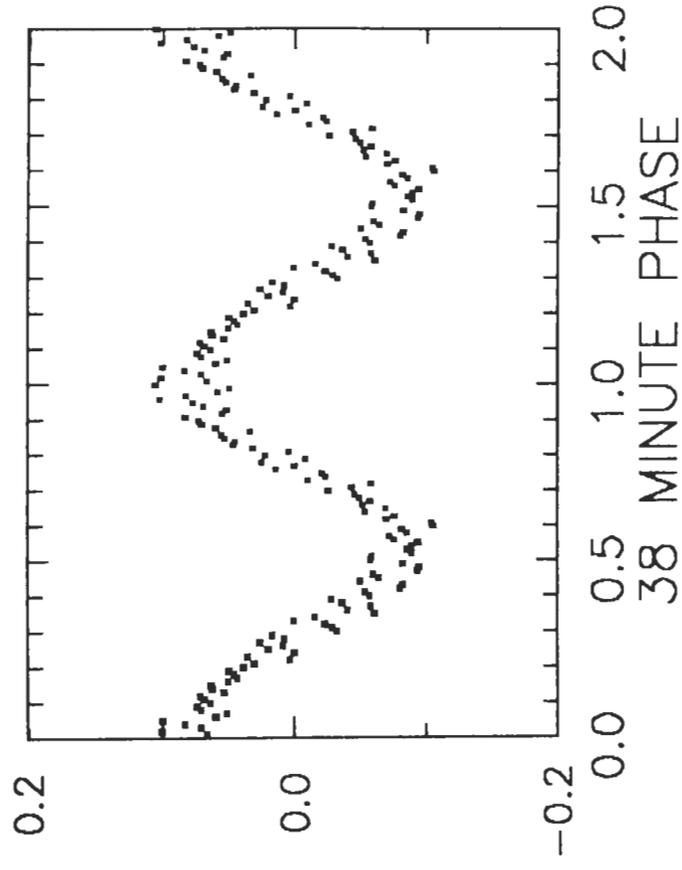
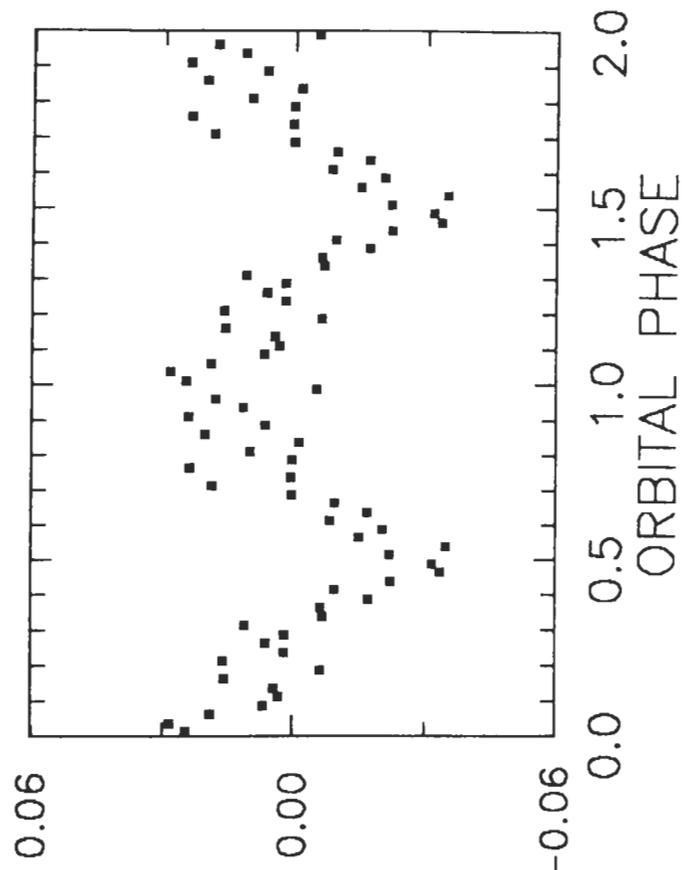
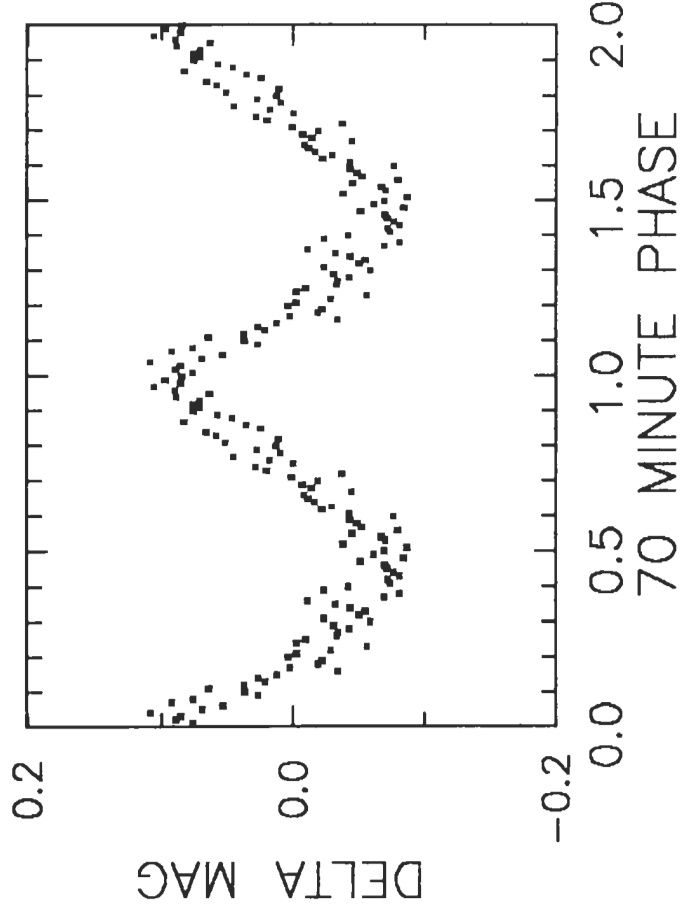
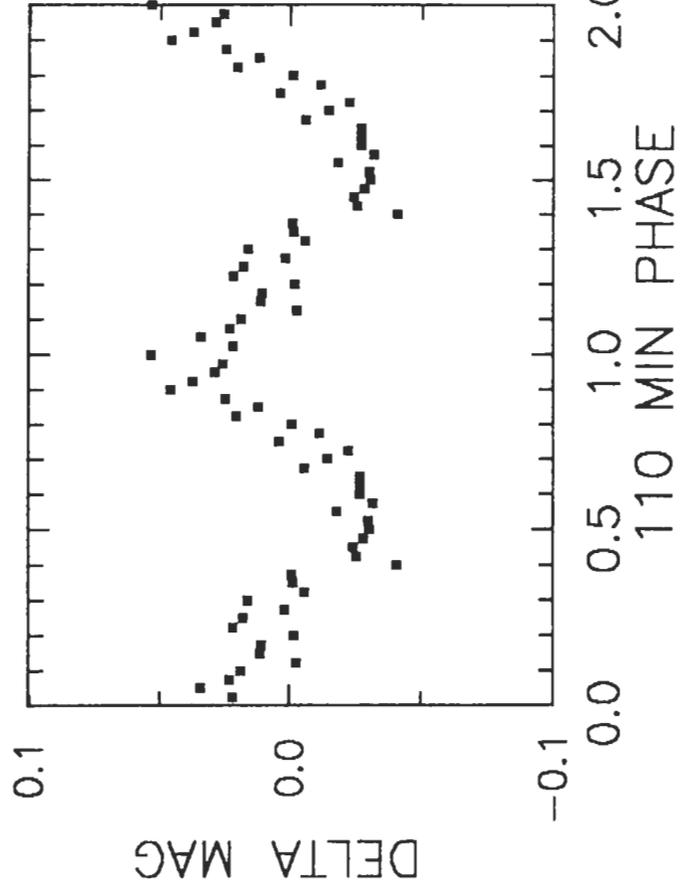

fig 5

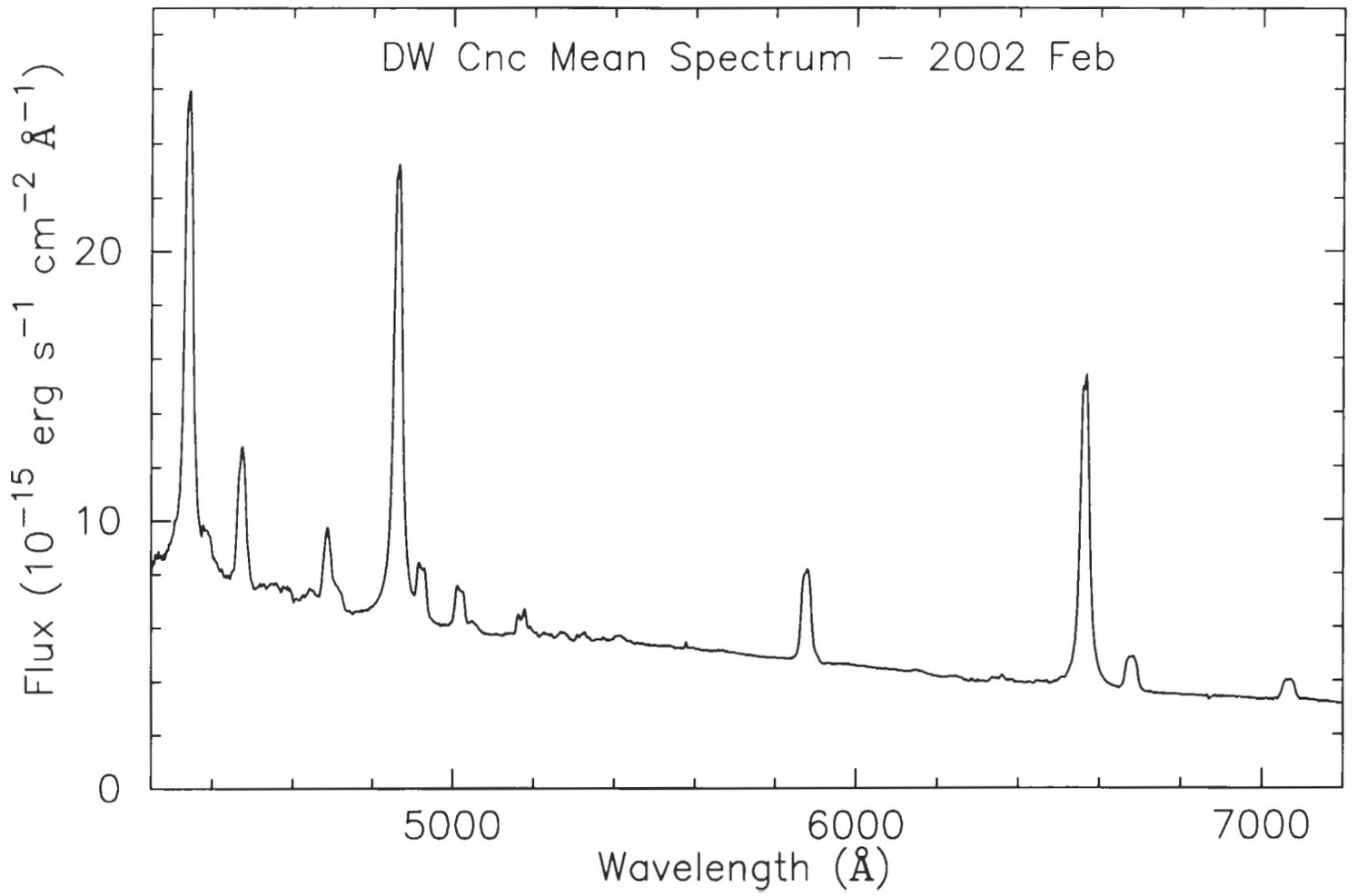

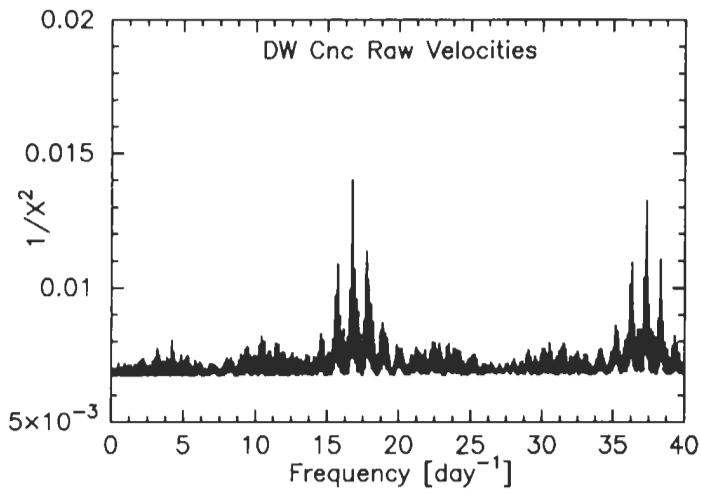
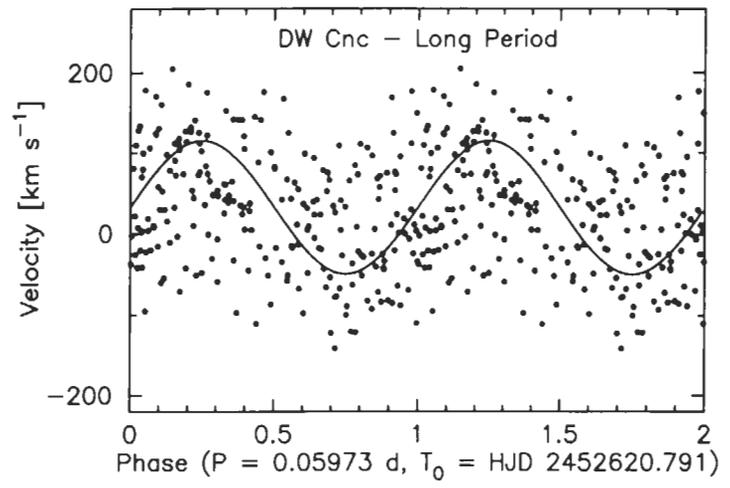
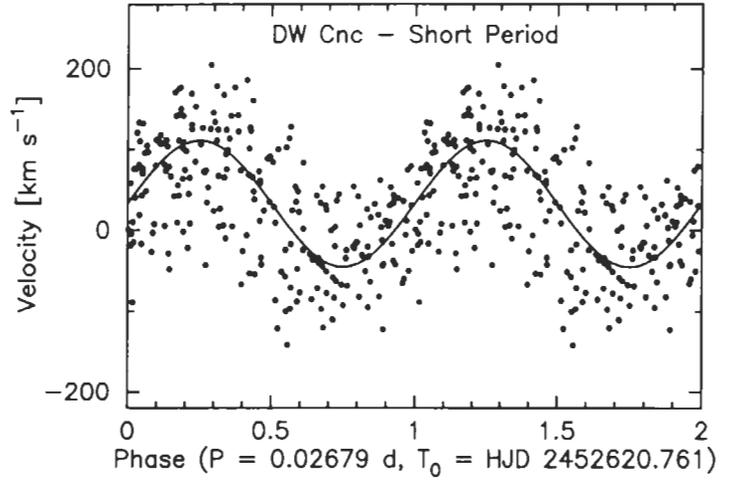
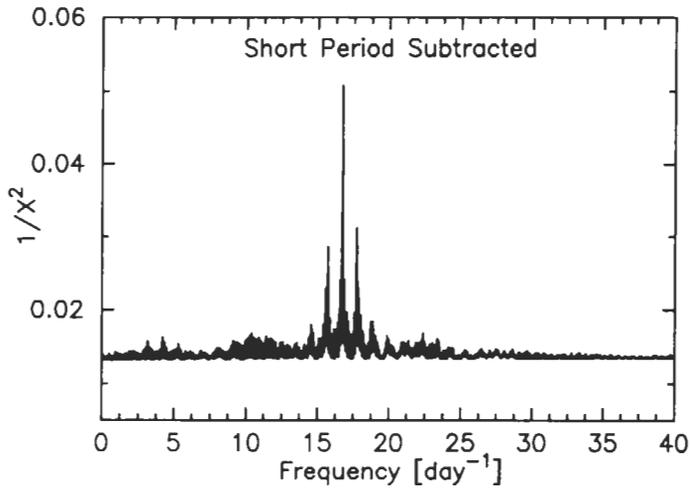
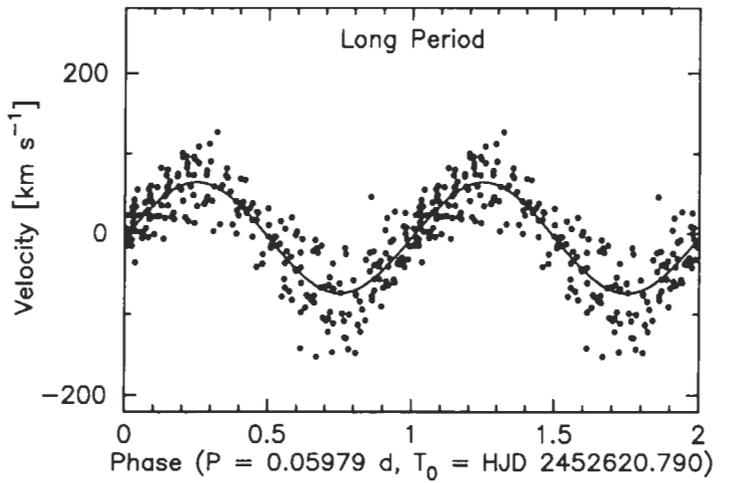
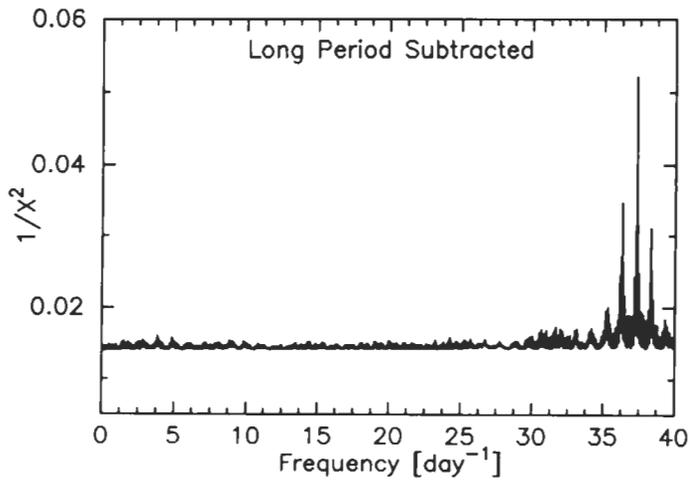
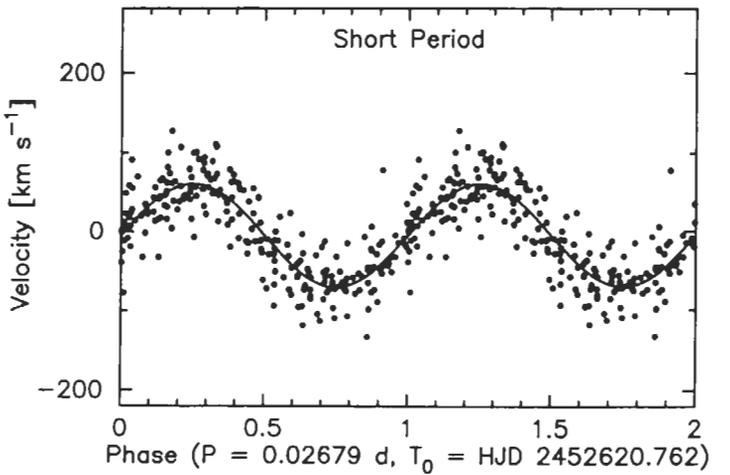

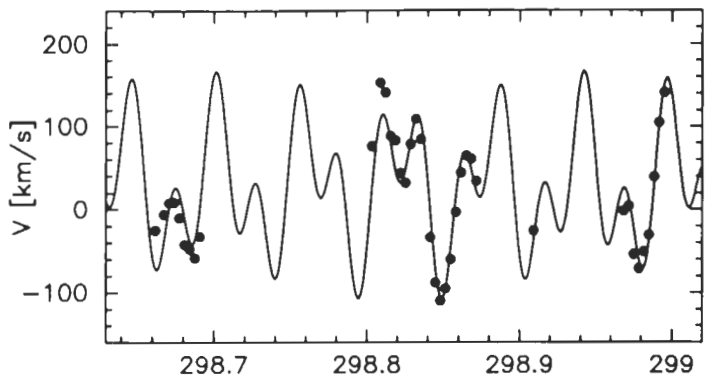
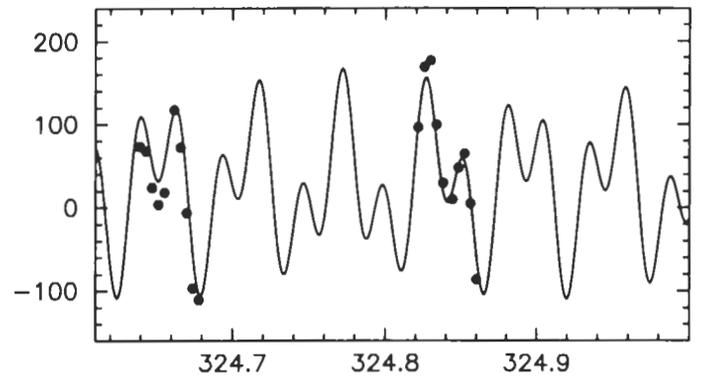
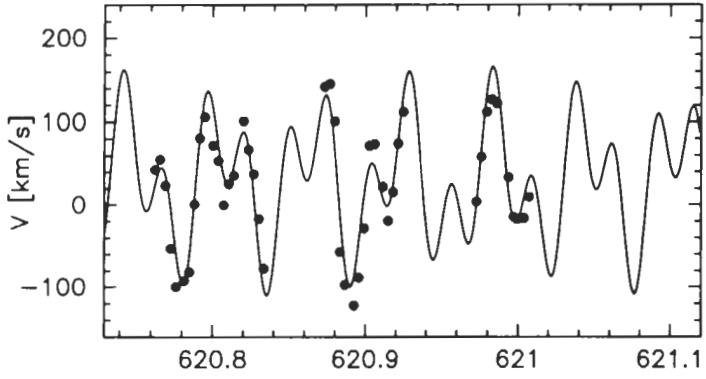
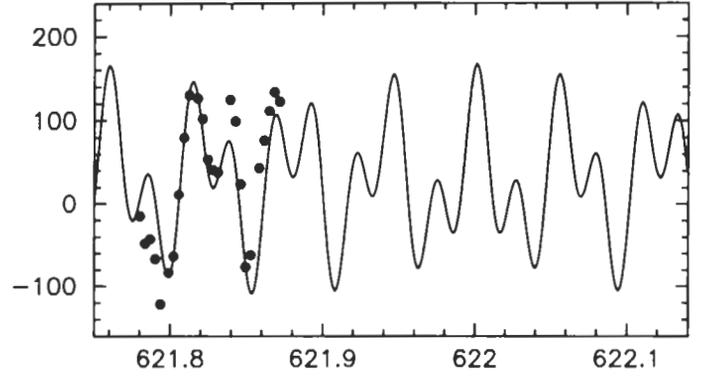
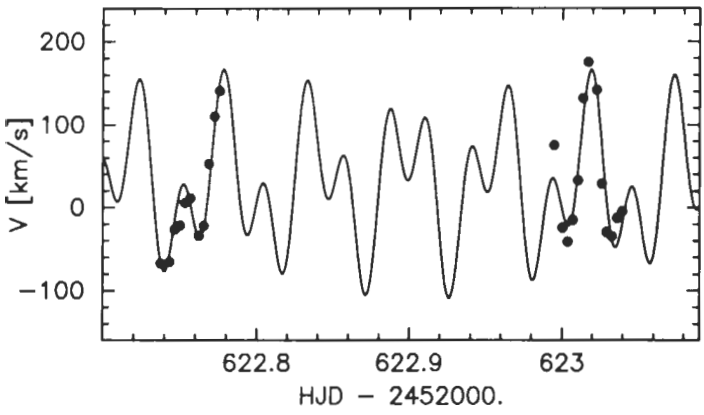
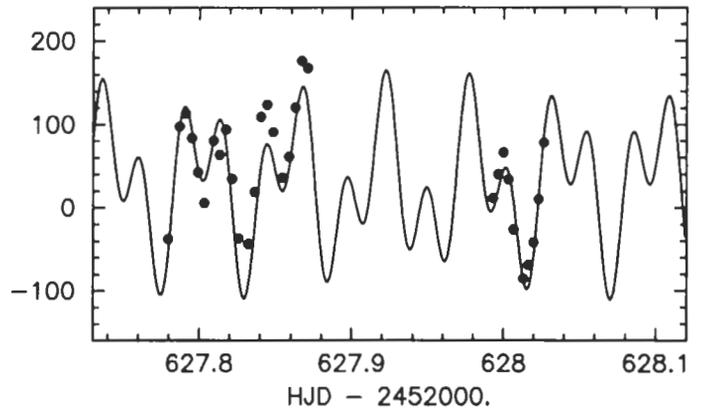

## DW Cnc – Phase-Averaged Spectra

Long Period

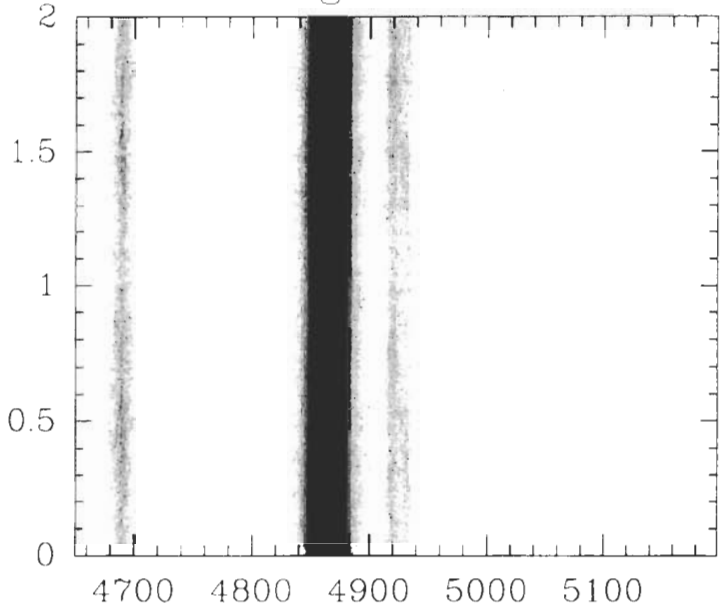
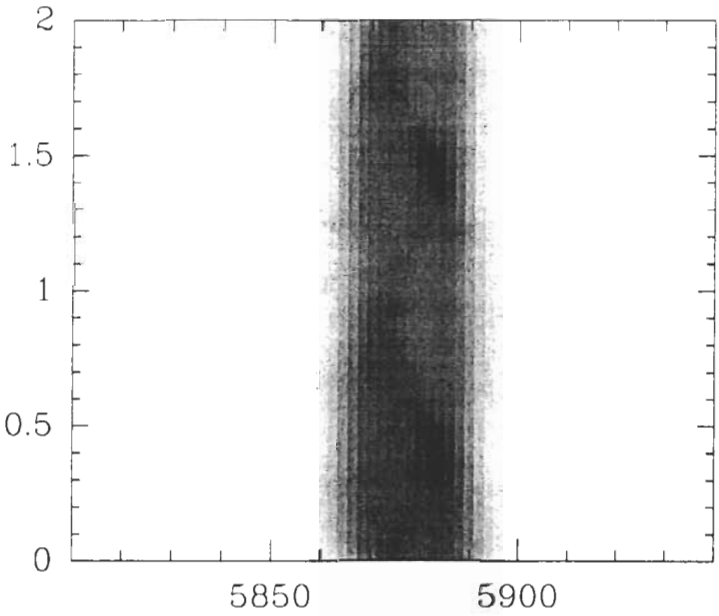
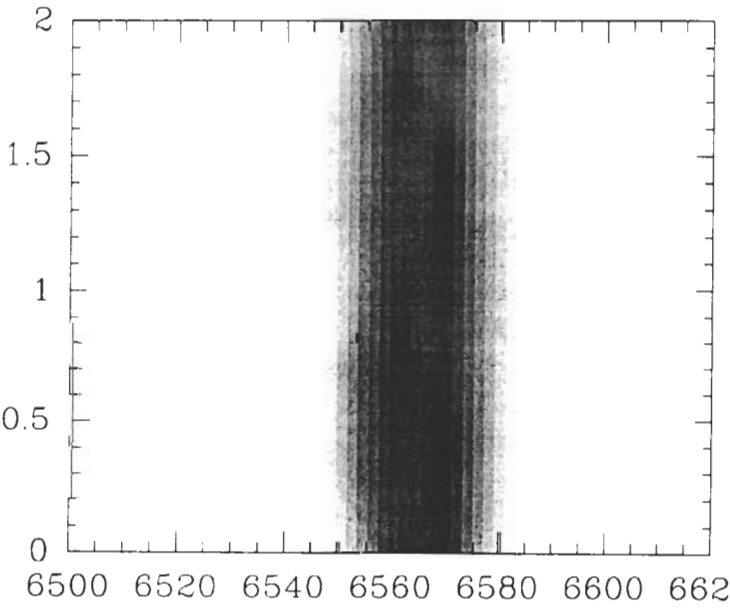

Short Period

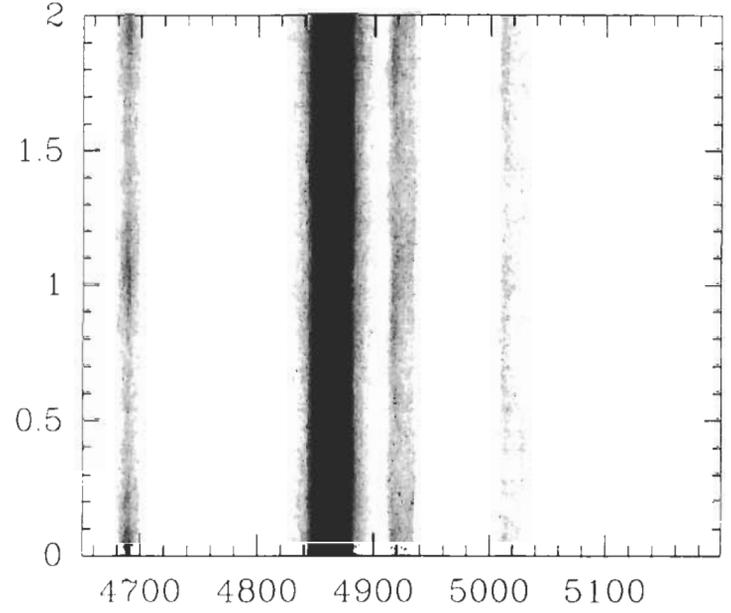
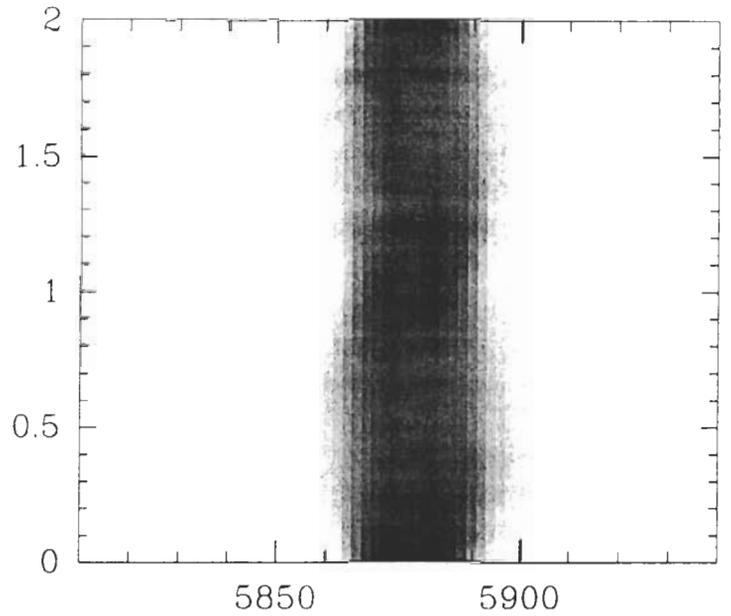
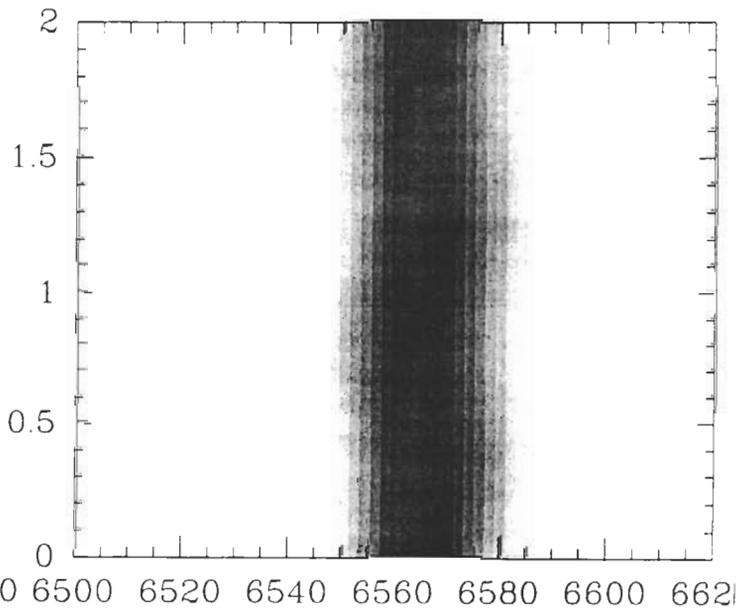